\documentclass{emulateapj}
\slugcomment{Accepted to  {\it The Astronomical Journal}}
\newcommand{\hii}{H{\small II} }

\def\farcm{\hbox{$~\mkern-4mu^{\prime}$}}
\def\farcs{\hbox{$~\mkern-4mu^{\prime\prime}$}}

\def\la{\mathrel{\hbox{\rlap{\hbox{\lower4pt\hbox{$\sim$}}}\hbox{$<$}}}}
\def\ga{\mathrel{\hbox{\rlap{\hbox{\lower4pt\hbox{$\sim$}}}\hbox{$>$}}}}

\shortauthors{Schenck}
\shorttitle{LMC ISM}

\begin{document}

\title{A Chandra Study of the Interstellar Metallicity in the Large Magellanic Cloud Using Supernova Remnants}

\author{Andrew Schenck, Sangwook Park, and Seth Post\\ Department of Physics, University of Texas at Arlington, Arlington, TX 76019; andrew.schenck@mavs.uta.edu; s.park@uta.edu; seth.post@mavs.uta.edu}


\begin{abstract}
We report on the results from our measurements of the interstellar medium (ISM) abundances for the elements O, Ne, Mg, Si, and Fe in the Large Magellanic Cloud (LMC). We used the archival {\it Chandra} data for sixteen supernova remnants (SNRs) in the LMC (0453--68.5, DEM L71, N23, 0519--69.0, N49B, N132D, N49, N206, 0534--69.9, DEM L238, N63A, Honeycomb, N157B, 0540--69.3, DEM L316B, and 0548--70.4). Our results represent LMC abundance measurements based on the modern {\it Chandra} data. We place tight constraints on our measured elemental abundances and find lower abundances than previous measurements by Hughes et al. (1998) (by a factor of $\sim$2 on average except for Si) who utilized similar methods based on a smaller sample of ASCA data of SNRs in the LMC. We discuss origins of the discrepancy between our {\it Chandra} and the previous ASCA measurements. We also discuss our results in comparisons with the LMC abundance measurements in literatures.

\end{abstract}


\section {\label{sec:intro} INTRODUCTION}
Measurements of the interstellar medium composition are essential for understanding galactic evolution and star-formation history. The Large Magellanic Cloud (LMC) provides an excellent laboratory for the study of its interstellar medium (ISM) thanks to its proximity ($\sim$50 kpc) and its low Galactic foreground absorption ($N_{H,Gal}\sim 6 \times$ 10$^{20}$ cm$^{-2}$) (Dickey \& Lockman 1990). Supernova Remnants (SNRs) are excellent probes of their surrounding ISM. The forward shock propagates through the ISM sweeping and heating it up. In young SNRs this swept-up ISM is hot enough to radiate in X-rays, allowing us to study interstellar metallicity based on X-ray observations. X-ray spectroscopy of SNRs to measure the metal abundances of gas-phase ISM was pioneered by Hughes et. al 1998 (hereafter H98) who utilized the integrated X-ray spectra of seven LMC SNRs, taken by ASCA. This method is independent of and complementary to the optical spectroscopy of stars and \hii regions (e.g., Russell \& Dopita 1992). With the high resolution imaging of modern {\it Chandra} data, we are now able to resolve shocked ISM features from regions with metal-rich ejecta thereby allowing us to perform accurate measurements of uncontaminated swept-up ISM elemental abundances (see Hughes et al. 2003, 2006; Park et al. 2003a, 2012;  Borkowski et al. 2007). Thus, revisiting these LMC abundance measurements based on the high resolution {\it Chandra} data of LMC SNRs is warranted. 

In this paper we analyze archival {\it Chandra} data of sixteen SNRs located throughout the LMC to estimate the ISM abundances of the LMC. Taking advantage of these high-resolution {\it Chandra} data (angular resolution of $\sim$0.5\farcs on axis, which is two orders of magnitude improvement from the ASCA data), we are able to effectively isolate the swept-up ISM from any metal-rich ejecta and/or pulsar wind nebulae (PWN) to make more realistic measurements of metal abundances in the LMC ISM. With our larger sample size we can also significantly reduce statistical uncertainties on the abundance measurements to help improve the accuracy of the LMC metallicity.

Here we report the results of our measurements of the ISM abundances of O, Ne, Mg, Si, and Fe in the LMC. In Section 2 we describe our data. In Section 3 we explain our regional selection, spectral extraction, and model fitting applied to each SNR. In Section 4 we compare this work with H98 results, comment on the utility of our new abundance measurements, explore possible spatial variations of the ISM composition throughout the LMC, compare this work to other abundance measurements, and briefly discuss individual SNRs. Finally, a summary is presented in Section 5.

\section{\label{sec:obs} OBSERVATIONS \& DATA REDUCTION}
There are 59 detected SNRs in the LMC (Maggi et al. 2015) and so far 25 of them are available in the {\it Chandra} archive. We selected sixteen SNRs based on two criteria: (1) they show a well-defined swept-up ISM shell, and (2) they are relatively bright and/or observed with a long exposure time to allow us statistically significant measurements of elemental abundances in its swept-up shell. All of these sixteen SNRs were observed using the ACIS-S3. The archival {\it Chandra} data of these sixteen SNRs are comprised of thirty-four ObsIDs (Table 1). Although there are a number of {\it Chandra} observations for SNR 1987A we chose to exclude this remnant in our study. X-ray emission from SNR 1987A is dominated by shocked circumstellar material from the massive progenitor (Sonneborn et al. 1987), which may not represent the ambient LMC abundances. We processed these individual ObsIDs using CIAO 4.3 and CALDB 4.4 which include corrections for the charge transfer inefficiency and the time-dependent contamination rate of the optical blocking filter. The change in the contamination rate of the optical blocking filter in 2009 reduced the quantum efficiency of the ACIS by an additional $\sim$15\% (which might have affected seven ObsIDs in our sample, in which the target SNRs were N49 and 0519-69.0). We then filtered these ACIS data following the standard data reduction procedure. There were periods of flaring background for observations of SNRs N63A and N49B. ObsID 777 (for the observation of N63A) had several short periods of high background ($\sim$2 ks in total), and ObsID 1041 (for the observation of N49B) had relatively high background for the last $\sim$4 ks out of the total $\sim$43 ks exposure. We removed these time periods from our analysis.

\section{\label{sec:result} ANALYSIS \& RESULTS}
Here we describe our general approach to measure ISM abundances for each SNR selected in this study. First we created a 3-color map for each SNR (with the color codes of red: 300--700 eV, green: 700--1100 eV, and blue: 1100-7000 eV) to locate the areas of soft X-ray emission (``red'' regions) presumably from the swept-up shell in its outermost boundary (Figure 1). We verified our candidate swept-up ISM regions for individual SNRs in literature (also see Section 4.4). We excluded regions for which X-ray emission from the shocked metal-rich ejecta has been reported in literature. Based on these criteria we selected several small and/or thin regions in the outermost boundary of each SNR (Figure 1). We defined individual regions to extract at least $\sim$3,000 counts (in the 0.3--7 keV band) in each region to allow statistically significant spectral model fits. We binned each spectrum to contain a minimum of 20 counts per energy channel. We fit these candidate ISM spectra using a nonequilibrium ionization (NEI) plane-parallel shock (vpshock in XSPEC) model \citep{bork01} with two foreground absorption column components, one for the Galactic column ($N_{H,Gal}$) and the other for the LMC column ($N_{H,LMC}$). The absorbing column for the Galaxy was fixed at $N_{H,Gal} = 6.5 \times 10^{20}$ cm$^{-2}$ for the direction toward the LMC \citep{DL90} while $N_{H,LMC}$ was allowed to vary. We used NEI version 2.0 (in XSPEC) associated with ATOMDB \citep{foster12} which includes inner shell lines and updated Fe-L lines (see Badenes et al. 2006). We performed background subtraction using source-free regions from just outside of each SNR. We fixed the redshift at {\it z} = 286 km $s^{-1}$ for the LMC (McConnachie 2012).

In this work we intend to measure abundances for all elemental abundances whose atomic line emission was present in the spectra. In the 0.3--7 keV band our sample of SNR spectra typically shows atomic line emission features for K-shell transitions in O, Ne, Mg, and Si as well as L-shell lines from Fe. Thus we fit the spectrum of each region in individual SNRs with the abundance values for O, Ne, Mg, Si, and Fe varied (hereafter, all elemental abundances are with respect to solar values [Anders and Grevesse 1989]). We found no significant variations of our estimated ISM abundances and electron temperature for the sub-regions within each individual SNR. We then averaged these measured abundance values for each SNR and summarize them in Table 2. We constrain the mean elemental abundances for each SNR within $\sim$20\% (uncertainties are with 90\% confidence, hereafter). Representing the mean interstellar abundances of the LMC, our average measured elemental abundances estimated from all sixteen SNRs, as listed in Table 2, are significantly lower (by a factor of $\sim$2 except for Si) than those measured by H98 (see Figure 3 for comparisons of the seven SNRs shared by both studies).

\section{\label{sec:disc} DISCUSSION}
\subsection{Comparisons with Previous ASCA Measurements}
H98 used ASCA data of seven bright LMC SNRs (N23, N49, N49B, N63A, DEM L71, N132D, and 0453-68.5) to estimate the elemental abundances of O, Ne, Mg, Si, S, and Fe in the LMC. Comparisons of the elemental abundances of the LMC ISM between this work and H98 are shown in Figure 3. Except for Si our measured elemental abundances are generally lower than H98 by a factor of $\sim$2. We identify several contributors to the discrepancy between H98 and our work. First, due to the large point spread function of the ASCA detectors ($\sim2\farcm$ half power diameter), H98 analyzed the X-ray spectra integrated over each entire SNR, while we used spectra extracted from resolved swept-up ISM features in each SNR. It has been well documented that N49, N49B, DEM L71 and N132D contain emission from metal-rich ejecta both in their central regions and in some cases even out near the outermost edges (Hughes et al. 2003; Park et al. 2003a, 2012; Borkowski et al. 2007). 

N49B contains Mg-rich ejecta in an extensive area in the SNR center (Park et al. 2003a). In fact, although statistical uncertainties are relatively large, H98 measured a higher Mg abundance ($\sim$50\% higher) for N49B than in other SNRs in their sample (see Figure 3). DEM L71 contains a Fe-rich central ejecta nebula (Hughes et al. 2003), and H98 measured a significantly higher Fe abundance in DEM L71 than in the other SNRs in their sample ($\sim$50\% higher, see Figure 3). N49 contains Si- and S-rich ejecta regions in its interior and in an ``ejecta bullet'' feature in the southwestern boundary (Park et al. 2012). N132D shows enhanced O abundance throughout its interior (Borkowski et al. 2007). We excluded these ejecta-dominated regions in our abundance measurements. Any embedded pulsar and/or PWN might also have affected abundance measurements in the ASCA data. N49 is coincident with an X-ray counterpart of a soft Gamma-ray repeater (SGR 0526-66; e.g., Park et al. 2012) while 0435-68.5 contains a PWN (e.g., McEntaffer et al. 2012). The continuum-dominated emission from these pulsars and/or PWNe might have resulted in underestimates of abundances when the integrated SNR spectrum was used to measure them.

To investigate the effect from metal-rich ejecta on ISM abundance measurements we have taken two regions from DEM L71 (one ISM region [the yellow region in Figure 4] and one metal-rich ejecta region [red region in Figure 4]) with similar photon statistics. We performed spectral model fits for the ISM region, the metal-rich ejecta region, and the combined ISM $+$ metal-rich ejecta region (Figure 4). We fixed in our measured ISM abundances for DEM L71 (see Table 2) into each spectral model, and allowed the electron temperature, ionization timescale, and normalization to vary. The fit was statistically acceptable for the ISM region ($\chi_{\nu}^2=1.01$). The fit was statistically unacceptable for the metal-rich ejecta region ($\chi_{\nu}^2 = 7.0$) and the fit was also statistically unacceptable for the combined spectrum ($\chi_{\nu}^2=2.85$). We then re-fit the three spectra allowing the elemental abundances for O, Ne, Mg, Si, and Fe to vary, and found statistically acceptable fits for all three spectra ($ISM\chi_{\nu}^2=1.0$, $Ejecta\chi_{\nu}^2=1.1$, $ISM+Ejecta\chi_{\nu}^2=1.1$). For the ISM-only spectra we found metal abundances similar (within $\sim$10\% uncertainties) to our measurements for DEM L71 shown in Table 2. For the metal-rich ejecta region we found significantly enhanced abundances compared to the ISM-only spectrum abundances: the Si and Fe abundances are higher than the ISM-only values by a factor of $\sim$2 and an order of magnitude, respectively. For the combined spectra, as perhaps expected, we estimated moderately enhanced abundances compared to the ISM-only abundances: i.e., the Ne and Fe abundances are higher than the ISM-only values by a factor of $\sim$2, and the Si abundances is $\sim$20\% higher than the ISM-only value. We note that these moderately enhanced abundances in the combined spectrum are in good agreement with those measured by H98 for DEM L71. Our test analysis of N49, N49B, and N132D shows similar results.

Similarly we show the effect of PWN on ISM abundance measurements. We selected one ISM region (the green region in Figure 5) and one region from the PWN (the red region in Figure 5) from 0453-68.5 with similar photon statistics. We performed spectral model fits for the ISM region, PWN region, and the combined ISM $+$ PWN region (Figure 5). We fixed in our measured ISM abundances for 0453-68.5 (Table 2) into each spectral model, and only allowed the electron temperature, ionization timescale, and normalization to vary. The fit was statistically acceptable for the ISM-only region ($\chi_{\nu}^2=1.1$ with energy range and E = $0.3-2.0$ keV). The fit was statistically unacceptable for the PWN region ($\chi_{\nu}^2=2.5$ with energy range E = $0.3-5.0$ keV) as well as the ISM$+$PWN region ($\chi_{\nu}^2=2.9$ with energy range E = $0.3-5.0$ keV). We then re-fit all spectra allowing the elemental abundances to vary. We found statistically acceptable fits for each individual spectrum ($ISM\chi_{\nu}^2 = 1.1$, $PWN\chi_{\nu}^2=1.3$, and $ISM_PWN\chi_{\nu}^2=1.4$). For the ISM-only region we measured metal abundances similar (within $\sim10\%$ uncertainties) to our measurements for 0453-68.5 shown in Table 2. For the PWN region we found somewhat low abundances ($\la$0.1 solar) compared to those in the ISM-only regions (Table 2). For the ISM$+$PWN region we also found somewhat lower abundances ($\la$0.1 solar) compared to those in the ISM-only regions (Table 2). H98 indeed found $\sim$10-50\% lower Ne, Mg, and Si abundances compared to the rest of the SNRs in their sample, and they interpreted that the lower mean abundance for this remnant is due to less contamination by metal-rich ejecta (see Figure 5 in H98). 


To test the model-dependence of abundance measurements between H98 and this work we fit the ASCA data of the seven SNRs (used in H98) with the NEI plane-shock model used in this work. Based on our plane-shock model fits of ASCA data, we measured the elemental abundances of O, Ne, Mg, Si, and Fe (Figure 6). Our abundance measurements for Ne and Fe are consistent with those by H98, while our O, Mg, and Si abundances are somewhat different than those by H98 (Figure 6). This suggests a model-dependence in abundance estimates for O, Mg and Si. Also, we fit the integrated {\it Chandra} spectra (extracted from each of the seven SNRs) with our NEI plane-shock model, and measured the elemental abundances of O, Ne, Mg, Si, and Fe (Figure 6). Abundance measurements are generally consistent between these {\it Chandra} and ASCA plane-shock model fits, whereas the Fe abundance is somewhat lower in our {\it Chandra} measurement. These results suggest that, in addition to the ejecta effect, there appears to be a model-dependence in abundance measurements as well as some systematic uncertainties between {\it Chandra} ACIS and ASCA SIS data (probably at a less-significant level).

Finally, we note that the {\it Chandra} ACIS-S3 detector is significantly more sensitive than that of the ASCA SIS detectors with a $\sim$three times larger effective area at E $\la$ 1 keV. Considering that O, Ne, and some Fe-L emission lines are located at energies below 1 keV, measuring abundances for those elements based on {\it Chandra} data would be generally more advantageous.

\subsection{Application of Our New LMC Abundances: Case of N63A}
X-ray emission from some LMC SNRs has been shown to be dominated by emission from the swept-up ISM, because their measured elemental abundances are roughly consistent with the previously known LMC abundances (e.g., N132D [Borkowski et al. 2007], N63A [Warren et al. 2003], etc). Based on our new measurements of the LMC abundances, we revisited some of those ISM-dominated LMC SNRs. Here we briefly discuss the case of N63A as an example. We divided N63A into many small sub-regions utilizing an adaptive mesh grid. Each region in the grid was created to contain a minimum of 3,000 photon counts in the 0.3--7 keV band. Our adaptive mesh created 346 regions for N63A (Figure 7). We extracted the spectrum from each region. Then we performed two different spectral model fits for each regional spectrum utilizing a NEI plane-shock model; one fit with elemental abundances fixed at our mean LMC ISM values (Table 2), and the other one with abundances fixed at H98's mean abundance values (Table 2). We allowed the foreground absorption column, electron temperature, ionization timescale, and normalization to vary. From these fits we created a map of the reduced chi-square ($\chi_{\nu}^2$) for each of these two model fits for all 346 regions (Figure 7). The overall $\chi_{\nu}^2$ values are significantly smaller in the model fits with our new LMC abundances than those with H98 abundances (e.g., 243 out of 346 regions show $\chi_{\nu}^2<1.6$ with our new LMC abundances [Figure 7a] while only 88 out of 346 regions show similarly good fits with the H98 abundances [Figure 7b]). In these model fits H98 abundances typically overestimate the observed atomic line fluxes. 

We note that some sub-regions of N63A show a high $\chi_{\nu}^2$ when their spectra are fitted with our new LMC abundances (particularly around the central parts of the SNR). For these regions our LMC abundances under-predict the observed emission line fluxes, suggesting the presence of overabundant ejecta-dominated regions there. It is remarkable to reveal the (previously-unknown) candidate ejecta dominated regions in N63A based on our new LMC abundance values. The detailed study of N63A is beyond the scope of this work, and we will present our detailed analysis of N63A in a separate paper. These results from our case study of N63A show the utility of measuring the accurate values of LMC ISM abundances.

\subsection{Comments on Spatial Variation of LMC Abundances}
We explore variations of the ISM abundances across the LMC (see Figure 2 for each SNR's location across the face of the LMC). We first investigate any spatial variation of our measured LMC abundances depending on SNR types (Type Ia vs core-collapse (CC)) with which we measured their ``local'' ISM abundances. In our sample we have ten CC SNRs, five Type Ia SNRs, and one un-typed SNR (Table 2). We find no significant variation in elemental abundances for Ne, Mg, Si, or Fe among these different types of SNRs (Figure 8). The O abundance appears to show a marginal enhancement for N63A, N49, and N49B (the best-fit values are $\sim$50\% higher than those for the rest of the SNRs in our sample). All of these three SNRs are located near the northern boundary of the LMC (Figure 2). Possibly, this part of the LMC may contain O-enhanced environments than the rest of the LMC. 

We also explore variations of ISM abundances between star-forming (SF) and non-star-forming (NSF) environments. In our sample there are ten SNRs located in star-forming environments while the remaining six SNRs are located in non-star-forming environments (see Figure 2). For Ne, Mg, and Si there are no significant variations between SNRs in either environment. The O abundance shows marginal evidence of enhancements in star-forming environments (Figure 9). This is likely due to the moderate enhancement of the O abundance in N63A, N49, and N49B (Figure 8), all of which are located in star-forming environments near the northern boundary of the LMC. We note that the 30 Dor region (the largest active star-forming region in the LMC) does not show evidence for an O-enhancement compared to the other parts of the LMC. Thus, if the O enhancement in the northern boundary of the LMC is true, its origin is unclear. It is also worth mentioning that the Fe abundance appears to be marginally enhanced in the non-star-forming regions. However, this suggestive Fe-enhancement is not conclusive in the current data, and thus it is difficult for us to further discuss it.

\subsection{Comparisons With Other LMC Abundance Measurements}
Recent optical measurements of elemental abundances in the LMC use large samples of field stars (Cole et al. 2005; Pompéia et al. 2008; Lapenna et al. 2012; Van der Swaelmen et al. 2013) to measure the elemental abundances of O, Mg, Si, and Fe as well as much heavier elements. Lapenna et al. (2012)'s O, Mg, and Si abundances are all consistent with our abundance measurements (Figure 10). Lapenna et al. (2012) showed that metallicity distributions tend to peak at about -0.57 dex for most field star samples; e.g., $[$Fe/H$]$ = 6.9. This is in general agreement (within $\sim$10\%) with our measurement of $[$Fe/H$]$ = $6.84^{+0.03}_{-0.02}$. Also, our results are in good agreement with those by Korn et al. (2000) for Mg and Si (Figure 10). There is discrepancy between our LMC abundance measurements and those by Russell and Dopita (1992): i.e., our estimates are generally lower by a factor of $\sim$2-3. The origins of this discrepancy are unclear. Possibly there are systematic differences between measuring the ISM abundances using different objects (e.g., SNRs vs \hii regions/stars). 

Recently Maggi et al. (2015) performed an X-ray study of 51 SNRs in the LMC utilizing {\it XMM-Newton} data. They detected metal-rich ejecta in a large portion of their sample (39 out of 51 SNRs). The remaining twelve SNRs were assumed to be dominated by swept-up ISM emission. They used the integrated spectrum from each individual SNR to measure the ISM abundance of the LMC. To increase their sample size they included SNRs (among the ejecta-dominated SNRs) that showed ISM-like abundances for at least one element. Then, they measured the LMC abundances for O, Ne, Mg, Si, and Fe. Their measurements ($[$O/H$]$ = $8.01^{+0.14}_{-0.21}$, $[$Ne/H$]$ = $7.39^{+0.11}_{-0.15}$, $[$Mg/H$]$ = $6.92^{+0.20}_{-0.37}$, $[$Si/H$]$ = $7.11^{+0.20}_{-0.40}$, and $[$Fe/H$]$ = $6.97^{+0.13}_{-0.18}$, where $[$X/H$]$ = 12 + log(X/H)) are generally consistent with our measured elemental abundances (Figure 10). We note that Maggi et al's results are based on the integrated spectrum from each SNR (due to the poor resolution of the EPIC detectors) rather than the spatially-resolved spectroscopy used in our study. Also, statistical uncertainties on their ISM abundance estimates are significantly larger than those on our measurements typically by a factor of $\sim$3-4. While LMC abundance measurements are generally consistent between Maggie et al. (2015) and this work, we conclude that our results based on the high resolution {\it Chandra} data represent more accurate estimates of the LMC abundances.

\subsection{Summary for Individual SNRs}
\subsubsection{0453-68.5}
0453-68.5 is a faint, aged ($>$17,000 yr), SNR showing a clear limb-brightened shell that is dominated by X-ray emission from the swept-up ISM (McEntaffer et al. 2012). 0453-68.5 contains a PWN at its center solidifying its classification for a CC explosion (Gaensler et al. 2003; McEntaffer et al. 2012). We selected six regions from around the outer swept-up shell, as shown in Figure 1a, to estimate abundances for O, Ne, Mg, and Fe. Our abundance measurements from these six sub-regions are all consistent within statistical uncertainties, and thus we summarize the average abundances in Table 2. We were unable to constrain the Si and S abundances in 0453-68.5 because of the spectrally-soft nature of the X-ray emission (photon statistics are poor at E $ > 1.5$ keV). McEntaffer et al. (2012) studied several regions partially overlapping with the regions used in this work. Their measured O abundance (O $= 0.17 \pm 0.02$) is significantly lower compared to Russell \& Dopita (1992) value (O $= 0.263$) but is consistent with H98's measurement within uncertainties (O = $0.21\pm0.05$). Our measured value for the O abundance (O $= 0.08\pm0.01$) is lower than that by H98. We note that McEntaffer et al. (2012) allowed only the O abundance to vary in their spectral model fits while fixing all other abundances at the LMC values by Russell and Dopita (1992). Thus, their O abundance measurement might have been affected by those fixed abundance values for other elements.
\subsubsection{DEM L71}
DEM L71 shows a double-shock morphology consisting of an outer swept-up ISM shell surrounding a central bright region of reverse-shock heated ejecta (Hughes et al. 2003). The composition of the metal-rich ejecta shows enhanced Si and Fe abundances, with a lack of O, classifying it as a Type Ia SNR (Hughes et al. 2003). We measured the elemental abundances for O, Ne, Mg, Si, and Fe based on spectra extracted from seven regions along the outer swept-up shell (Figure 1b). The measured elemental abundances were consistent among all these seven sub-regions, and the averaged abundance values are shown in Table 2. Our measured elemental abundances are generally in agreement with those found in inner rim region of the swept-up shell by Hughes et al. (2003).
\subsubsection{N23}
N23 is an irregular shaped SNR which appears to originate from a CC explosion of a massive star (Hughes et al. 2006). N23 shows a compact object near its center. This compact source is probably the stellar remnant from the explosion of a massive star, although its association with the SNR is still under debate (Hughes et al. 2006, Hayato et al. 2006). Hughes et al. (2006) found that most of the emission from N23 shows LMC-like elemental abundances, and found only one region (located near the center of the remnant, west of the compact object) with enhanced elemental abundances ($\sim$2-3 $\times$ LMC abundances). We selected four regions from all around the outer boundary of the SNR (Figure 1c, avoiding the metal-rich region found by Hughes et al. 2006) and measured the elemental abundances for O, Ne, Mg, Si and Fe. The measured elemental abundances are consistent among the four sub-regions. We present the average values for our measured abundances in Table 2. Our measured ISM abundance values are in good agreement with those in the east rim estimated by Hughes et al. (2006).
\subsubsection{0519-69.0}
0519-69.0 has been classified as a Type Ia SNR by Hughes et al. (1995). They found that 0519-69.0 is an O-poor SNR with Fe-rich ejecta throughout its central regions. 0519-69.0 shows a clumpy morphology with a clear inner ejecta nebula surrounded by a thin shell of swept-up ISM (Kosenko et al. 2010). We selected four thin red regions (dominated by spectrally soft X-ray emission) along the outermost boundary of the SNR (Figure 1d) to measure the elemental abundances for O, Ne, Mg, Si, and Fe. The measured elemental abundances are consistent among all four sub-regions. The averaged abundances are shown in Table 2. Abundance measurements from this outer swept-up shell of 0519-69.0 have not been previously reported in literature.
\subsubsection{N49B}
N49B shows Mg- and Si-rich ejecta near its center (Park et al. 2003a). N49B shows bright circumferential filaments that are emission from dense shocked ISM. We selected six regions along the outermost boundary (Figure 1e). We measured the elemental abundances for O, Ne, Mg, Si and Fe. Our measured elemental abundances are consistent among all six sub-regions, and the average abundance values are shown in Table 2. Our measured abundances are in agreement with those measured in the bright southern swept-up shell by Park et al. (2003a).
\subsubsection{N132D}
N132D is the brightest SNR in the LMC and belongs to the rare class of O-rich SNR. Optical observations have shown high-velocity O-rich ejecta, indicating a CC origin for this SNR (Morse, Winkler, \& Kirshner 1995). N132D interacts with CO clouds located all around the SNR supporting its CC origin (Banas et al. 1997; Sano et al. 2015). N132D shows a complicated filamentary morphology in X-rays with moderately enhanced metal-rich ejecta in its central regions (Borkowski et al. 2007). We measured the elemental abundances of O, Ne, Mg, Si and Fe based on spectra extracted from eight sub-regions from the outermost boundary (Figure 1f) and bright central filaments of the shocked ISM. The measured abundances are consistent among these eight sub-regions, and we present average abundance values in Table 2. We note that in their ejecta spectral analysis Borkowski et al. (2007) assumed 0.4 solar abundances for the underlying emission component originating from the swept-up medium.
\subsubsection{N49}
A soft Gamma-ray repeater (SGR 0526-66), a rare class of highly-magnetized neutron stars, is projected within the boundary of N49 (Cline et al. 1982; Rothschild et al. 1994), suggesting a CC origin for N49, although the physical association between SGR 0526-66 and N49 is unclear (Gaensler et al. 2001; Klose et al. 2004; Badenes et al. 2009; Park et al. 2012). N49 shows some regions of Si- and S-rich ejecta (including an ejecta ``bullet'' in the southwest boundary [Park et al. 2012]). N49 is also interacting with clumpy molecular clouds (Vancura et al. 1992) in the eastern and southeastern regions giving rise to complicated spectral nature of X-ray emission in that area (Park et al. 2003b). Avoiding these ejecta features and regions where the shock is interacting with clumpy molecular clouds, we selected two regions from the outermost boundary in the southern rim of the SNR (Figure 1g). Our measured abundances are consistent between the two selected regions, and are also in agreement with those measured by Park et al. (2012). We measured the elemental abundances for O, Ne, Mg, Si and Fe, and the average abundance values are shown in Table 2. 
\subsubsection{N206}
N206 is a mixed-morphology SNR (Williams et al. 2005) with a bright eastern band surrounded by a diffuse swept-up ISM shell. N206 is suggested be to a CC SNR based on observed ejecta elemental abundance structure (Williams et al. 2005). We selected two regions from the diffuse ISM shell (Figure 1h) and measured the O, Ne, Mg, Si and Fe abundances. The elemental abundances were consistent between these two sub-regions. The average abundance values are shown in Table 2. We note that our measured abundances are lower (by a factor of $\sim$2 for O, Ne, Mg and Fe) than those estimated by Williams et al. (2005) while our Si value is similar to that estimated by Williams et al. (2005). This discrepancy appears to be due to metal-rich ejecta contamination of the ISM abundance measurements in Williams et al. (2005) from an ejecta feature in the western part of their ISM shell region. 
\subsubsection{0534-69.9 and 0548-70.4}
0534-69.9 and 0548-70.4 show a bright central metal-rich ejecta nebula surrounded by a limb-brightened swept-up ISM shell (Hendrick et al. 2003). Based on low O/Fe abundance ratios in the central ejecta regions of both SNRs Hendrick et al. (2003) classified them as Type Ia explosions. In these SNRs the swept-up ISM emission can be clearly separated from the central ejecta emission. We extracted X-ray spectra from two outer boundary regions of 0534-69.9 (Figure 1i) to measure the elemental abundances for O, Ne, Mg, Si and Fe. We selected four regions in the outermost shell for 0548-70.4 (Figure 1p) and measured the elemental abundances for O, Ne, Mg, and Fe. The measured elemental abundances are consistent in all sub-regions in each SNR. The averaged elemental abundances for these SNRs are shown in Table 2. There are no previous estimates of metal abundances in the swept-up shell in these SNRs in literature.
\subsubsection{DEM L238}
DEM L238 is characterized by a bright central emission surrounded by limb-brightened swept-up ISM emission to the north, east, and west. The bright central emission shows enhanced Fe abundance suggesting that DEM L238 is a Type Ia SNR (Borkowski et al. 2006a). We selected two regions from the outer shell to measure the elemental abundances for O, Ne, Mg and Fe (Figure 1j). The measured elemental abundances from both regions are consistent with each other. The average elemental abundances are shown in Table 2. There are no previous measurements of the abundances in the swept-up shell in DEM L238.
\subsubsection{N63A}
N63A is the second brightest remnant in the LMC and is embedded in an \hii region with its position coincident with the OB association NGC 2030 (Warren et al. 2003). This suggests a massive Population I progenitor for N63A (Chu \& Kennicutt 1988). N63A shows many interesting morphological features such as loops and plumes at the outer-most boundary of the SNR along with many filamentary features throughout the SNR. The loop and plume features appear to show LMC-like abundances (Warren et al. 2003). However, we exclude these regions from our ISM study since their origin (shocked ISM vs ejecta) is still unclear (Warren et al. 2003). Warren et al. (2003) showed that the column density toward N63A varies across the SNR. We selected seventeen regions in N63A from all around the outermost boundary (Figure 1k). These regions show similar column densities to one another (as shown in Warren et al. 2003). Initially we fitted these seventeen sub-regional spectra with the column density varied. The measured column densities are all consistent within statistical uncertainties, and thus we repeated our spectral model fits for all of these sub-regions with $N_{H,LMC}$ fixed at the mean value ($N_{H,LMC}=1.1$ $\times$ $10^{21}$ cm$^{-2}$). We measured the elemental abundances for O, Ne, Mg, Si and Fe (Table 2). Our elemental abundances are in good agreement with those found by Warren et al. (2003).
\subsubsection{Honeycomb Nebula}
The Honeycomb nebula is a peculiar SNR. Its name is due to its optical morphology consisting of over 10 loops (with sizes of $\sim$2-3 pc) created by a SN blast wave interacting with sheets of dense but porous interstellar gas (Chu et al. 1995). Honeycomb's overall morphology is unlike most SNRs who show a coherent shell-like structure. We probed Honeycomb for possible metal-rich ejecta and found no evidence of enhanced elemental abundances. We identified an X-ray point source located within the southwestern limb as a CXOXASSIST source (X053543.01-691817.5). We found that the X-ray spectrum of this point source can be equally fitted by several different models because of the poor photon statistics (240 counts in the 0.5-7 keV band): e.g., a power law (photon index $\Gamma$ $\sim$ 1.9, $N_{H,LMC}$ $\sim$ 7 $\times$ $10^{20}$ cm$^{-2}$ with $\chi_{\nu}^2$ = 0.7), a blackbody (kT $\sim$ 0.6 keV, $N_{H,LMC}$ $\sim$ 2 $\times$ $10^{20}$ cm$^{-2}$ with $\chi_{\nu}^2$ = 1.4), and a plane shock (kT $\sim$ 8 keV, $N_{H,LMC}$ $\sim$ 5 $\times$ $10^{20}$ cm$^{-2}$ with $\chi_{\nu}^2$ = 1.1). We note that the best-fit electron temperature for the plane shock model is extremely high and it is unlikely adequate for this point source. This suggests either a power law or blackbody model would be more appropriate to describe the observed X-ray spectrum of this source. The best-fit column for the blackbody model ($N_{H,LMC}$ $\sim$ 2 $\times$ $10^{20}$ cm$^{-2}$) favors this source's physical association with the Honeycomb nebula, while the column implied for the best-fit power law model ($N_{H,LMC}$ $\sim$ 7 $\times$ $10^{20}$ cm$^{-2}$) is considerably larger than that for the Honeycomb nebula ($N_{H,LMC}$ $\sim$ 2 $\times$ $10^{20}$ cm$^{-2}$). However, the uncertainties on the column measurements for this point source in these models are large due to poor photon count statistics. Deeper observations are required to reveal the true nature of this X-ray point source. We extracted two spectra from Honeycomb to measure the ISM abundances (Figure 1l). One region from the entire northeastern limb and one region from the entire southwestern limb (excluding the X-ray point source). We measured the elemental abundances for O, Ne, Mg, and Fe (Table 2). There are no previous ISM abundance measurements for this SNR in literature.
\subsubsection{N157B}
N157B contains one of the most energetic pulsars known (PSR J053747.39 -691020.2). This pulsar is surrounded by a bright non-thermal X-ray nebula which likely represents a toroidal pulsar wind terminal shock observed edge-on (Chen et al. 2006). While N157B shows metal-rich ejecta through its interior close to the pulsar, diffuse swept-up ISM emission extends significantly to the north (Chen et al. 2006). We used one region from the northeast to measure the elemental abundances of O, Ne, Mg and Fe (Figure 1m). Our measured elemental abundances are shown in Table 2. There are no previous abundance measurements for the swept-up shell in this SNR.
\subsubsection{0540-69.3}
0540-69.3 has been identified as one of a handful of O-rich SNR in the LMC, and it contains a bright pulsar and PWN which have been throughly studied at many different wavelengths (e.g., Middleditch \& Pennypacker 1985; Manchester et al. 1993a, 1993b; Gotthelf \& Wang 2000; Kaaret et al. 2001; Petre et al. 2007). These O-rich ejecta and central pulsar solidify its CC origin. 0540-69.3 contains non-thermal emission in the form of ``arcs'' in the eastern and western boundaries of the SNR, which are suggestive of efficient cosmic-ray acceleration in the SNR shock (Park et al. 2010). Park et al. (2010) detected candidate metal-rich ejecta regions in the southern part of the SNR. We extracted spectrum from two regions in the western boundary of the SNR avoiding these PWN, non-thermal arcs, and candidate ejecta regions (Figure 1n). We measured the elemental abundances for O, Ne, Mg, Si, and Fe. Our measured elemental abundances are consistent between both regions and the average values are shown in Table 2. Hwang et al. (2001) fixed elemental abundances at the LMC values by Russell and Dopita (1992) in their spectral model fit for their ambient medium region. However, their overall model fits are statistically poor ($\chi_{\nu}^2=1.7-2.5$). They noted that their measurements might have been contaminated by metal-rich ejecta.
\subsubsection{DEM L316B}
DEM L316B is a faint SNR that shows several central knots surrounded by diffuse shell-like emission. DEM L316B has been typed as a CC SNR based on the observed ejecta abundance ratios (Williams \& Chu 2005). Avoiding metal-rich ejecta features throughout DEM L316B's center we extracted the X-ray spectrum from a shell region surrounding the entire SNR (Figure 1o) to measure the ISM elemental abundances for O, Ne, Mg, Si and Fe. Our swept-up shell region is similar to Region 12 in Williams \& Chu (2005). We are unable to compare our elemental abundance values to those found by Williams \& Chu (2005) since they did not exclusively provide the abundance values for Region 12. We note that  Williams \& Chu (2005) did find significantly under-abundant Fe (Fe $\sim 0.10 \pm 0.03$) in some regions of DEM L316B, which is consistent with our measurement of the Fe abundances in this SNR.

\section{Summary}
Using the high resolution archival {\it Chandra} ACIS data of sixteen SNRs in the LMC we measure the elemental abundances of O, Ne, Mg, Si and Fe of the gas-phase ISM in the LMC in an unprecedented precision in X-rays. For our spectral extractions we avoid any contamination from metal-rich ejecta and/or PWN. We find lower elemental abundances (by a factor of $\sim$2 for all elements except for Si) for the ISM in the LMC compared to those found by H98 who similarly used SNRs in the LMC (based on ASCA data). We attribute this discrepancy to several effects such as the overabundant metal-rich ejecta contribution in the ASCA data and spectral model-dependence of the abundance measurements. There also appears to be some systematic uncertainties between {\it Chandra} and ASCA detectors.

We find no significant variation of ISM abundances within individual remnants or between SNR types. The O and Fe abundances appear to show marginal evidence for variation between star-forming and non star-formation environments. To confirm or dispute this suggested spatial variation of abundances, deeper observational data with a larger sample may be required. We compared our results with optical measurements of LMC abundances in literature and found agreements with recent results (e.g., Korn et al. 2000; Lapenna et al. 2012). Our LMC abundance measurements are generally consistent with recent {\it XMM-Newton} results (Maggi et al. 2015), although the {\it XMM-Newton} results involve larger uncertainties than those on our measurements.

\acknowledgments

We would like to thank J. Y. Seok for supplying the H$\alpha$ map of the LMC. We would also like to thank Dr. John P. Hughes for his insight and helpful comments. This work has been supported in part by {\it Chandra} grant AR0-11008A.

\clearpage

%

\begin{figure*}[]
\figurenum{1}
\centerline{\includegraphics[angle=0,width=\textwidth]{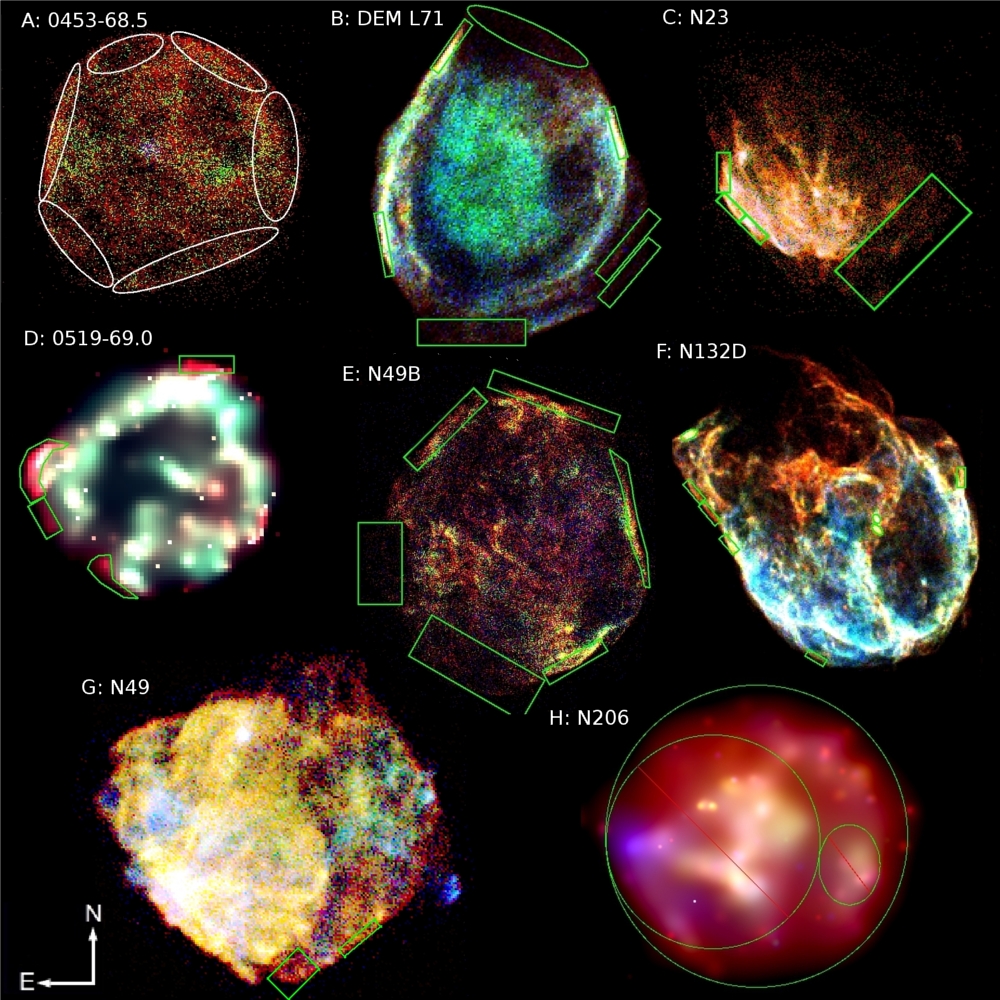}}
\figcaption[]{{3-color maps for individual SNRs. ISM regions used for spectral modeling are overlaid. Color codes are red: 300-720 eV, green: 720--1100 eV, blue: 1100--7000 eV. 0519-69.0 and N206 have been adaptively smoothed to emphasize their faint ISM features.}\label{fig:fig1}}
\end{figure*}

\begin{figure*}[]
\figurenum{1}
\centerline{\includegraphics[angle=0,width=\textwidth]{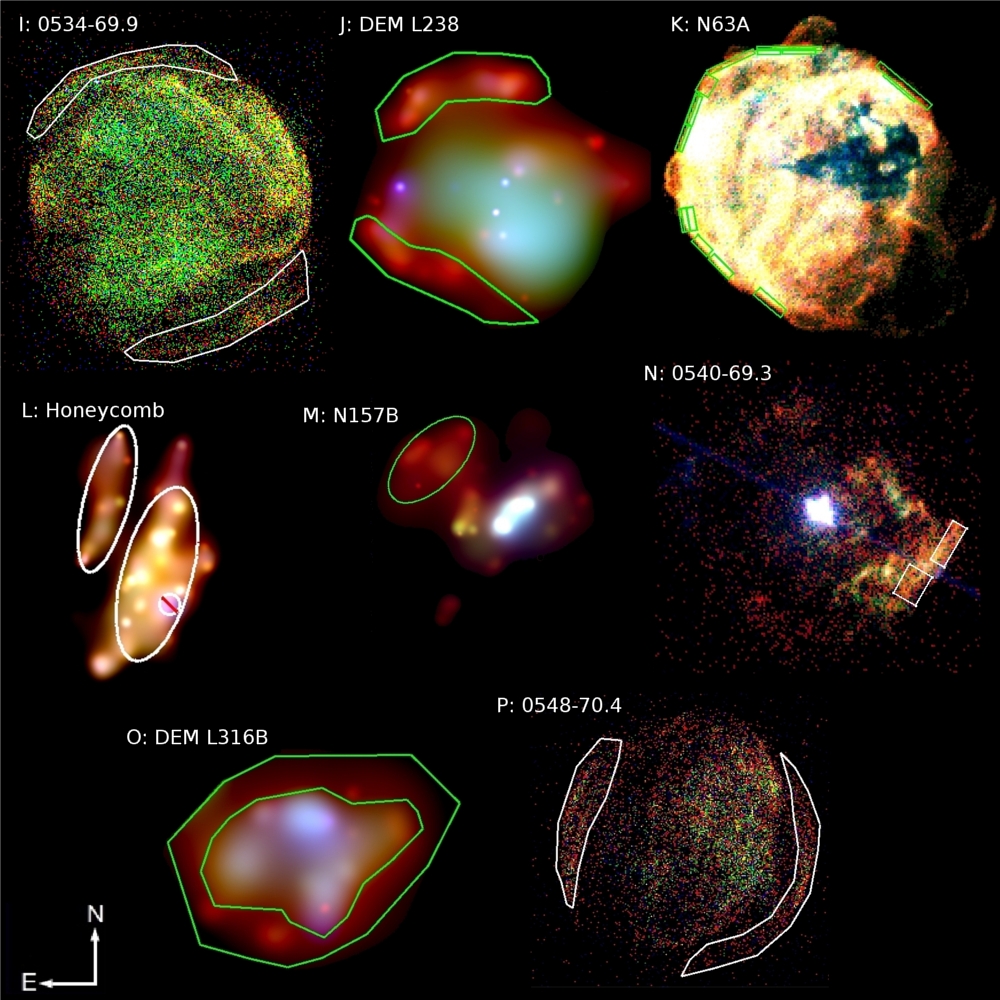}}
\figcaption[]{{3-color maps for individual SNRs with ISM regions used for spectral modeling overlaid. Color codes are red: 300-720 eV, green: 720--1100 eV, blue: 1100--7000 eV. DEM L238, Honeycomb, N157B, and DEM L316B have been adaptively smoothed to emphasize their faint ISM features.}\label{fig:fig1}}
\end{figure*}

\begin{figure*}[]
\figurenum{2}
\centerline{\includegraphics[angle=0,width=\textwidth]{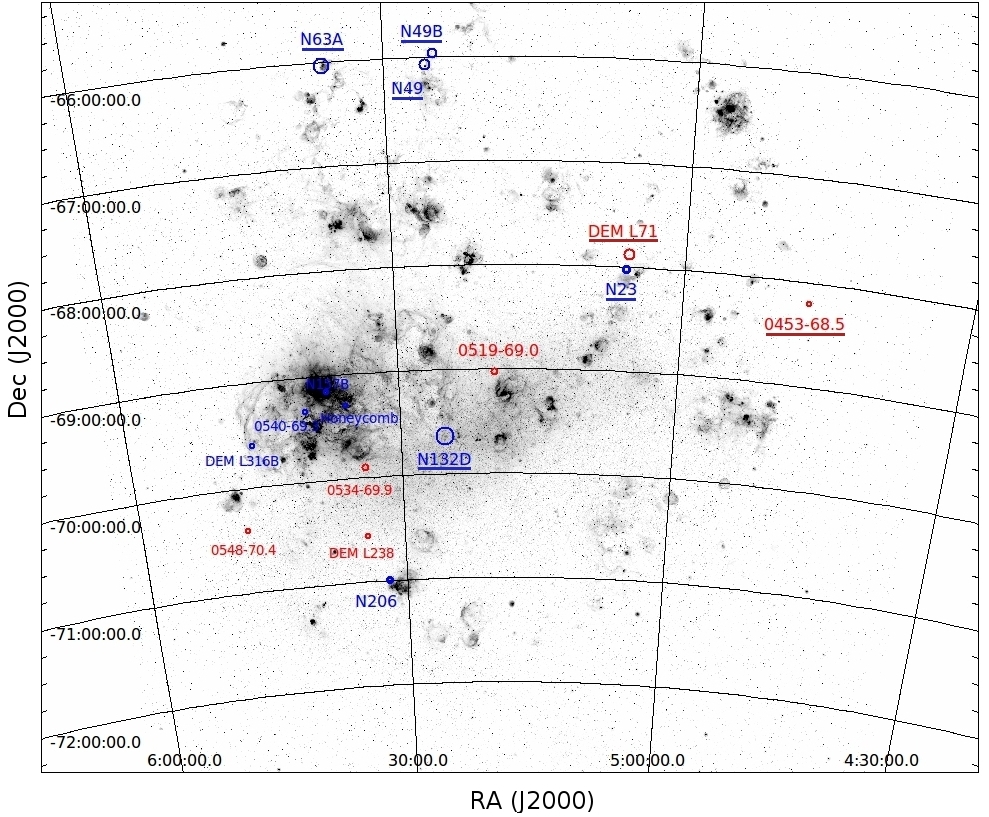}}
\figcaption[]{{The H$\alpha$ map of the LMC (Image Credit: C. Smith, S. Points, the MCELS Team and NOAO/AURA/NSF). Sixteen LMC SNRs used in this study are marked with circles. A larger circle implies a brighter SNR in a square-root scale. Blue circles indicate that the SNR is located within a star-formation region, while red circles indicate the SNR is not located within a star-formation region. The SNRs used in H98 are underlined.}\label{fig:fig2}}
\end{figure*}

\begin{figure*}[]
\figurenum{3}
\centerline{\includegraphics[angle=0,height=0.9\textheight]{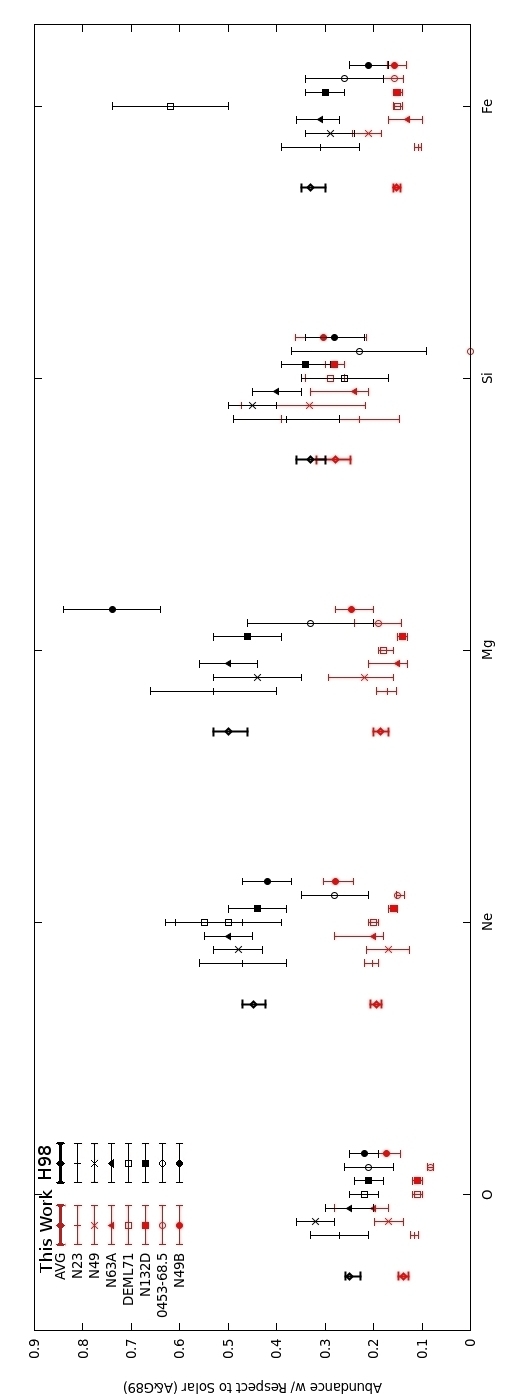}}
\figcaption[]{{Abundance comparisons between H98 and this work for seven SNRs used in both works. The left-most data point for each element is the mean value for the seven SNRs.}\label{fig:fig3}}
\end{figure*}

\begin{figure*}[]
\figurenum{4}
\centerline{\includegraphics[angle=0,width=\textwidth]{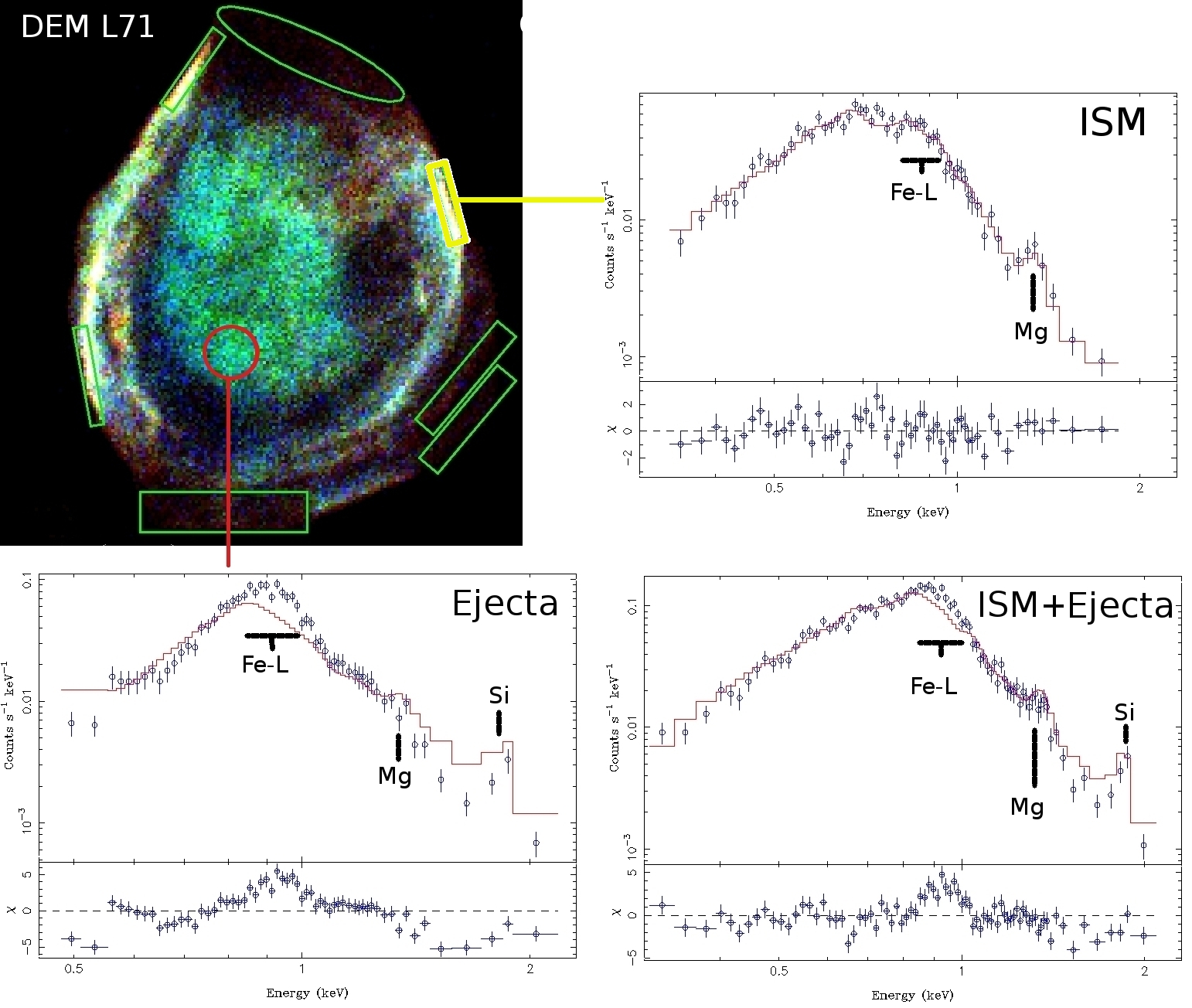}}
\figcaption[]{{Spectral model fits for three representative regions in DEM L71. In all fits metal abundances are fixed at our measured LMC abundance values (Table 2). ISM-only (yellow region), metal-rich ejecta only (red region), and their combined spectra. The residuals from the spectral model fits are shown at the bottom panel of each spectral plot. Green regions are the other regions used in our ISM abundance measurements.}\label{fig:fig4}}
\end{figure*}

\begin{figure*}[]
\figurenum{5}
\centerline{\includegraphics[angle=0,width=\textwidth]{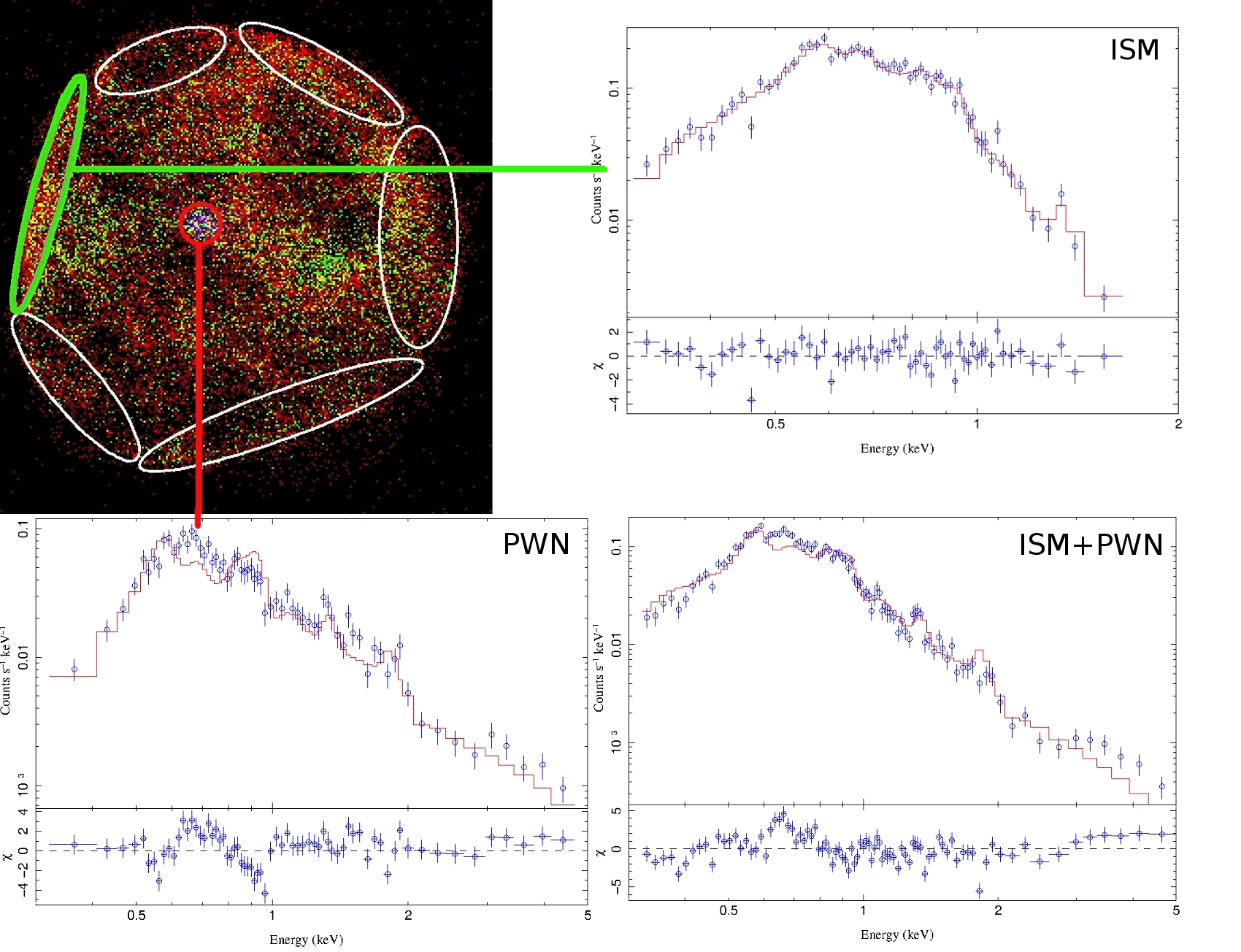}}
\figcaption[]{{Spectral model fits for three representative regions in 0435-68.5. In all fits metal abundances are fixed at our measured LMC abundance values (Table 2). ISM-only (green region), PWN only (red region), and their combined spectra. The residuals from the spectral model fits are shown at the bottom panel of each spectral plot. White regions are the other regions used in our ISM abundance measurements.}\label{fig:fig5}}
\end{figure*}

\begin{figure*}[]
\figurenum{6}
\centerline{\includegraphics[angle=0,width=\textwidth]{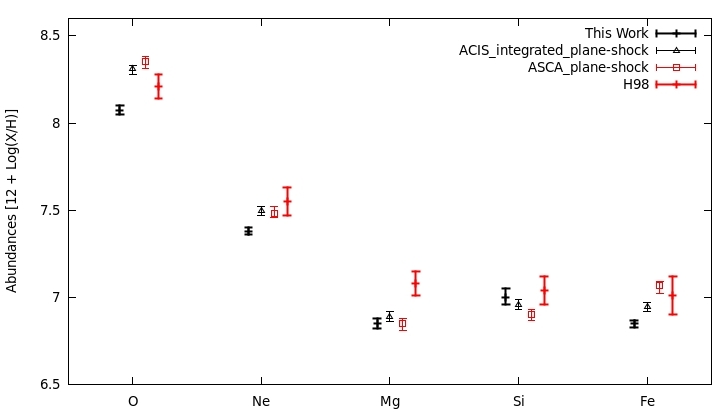}}
\figcaption[]{{Averaged elemental abundances estimated by our plane-shock model fits to integrated ACIS and ASCA spectra of the seven SNRs used in H98. Elemental abundances from this work and H98 are also shown. Statistical uncertainties are with 90\% confidence level.}\label{fig:fig6}}
\end{figure*}

\begin{figure*}[]
\figurenum{7}
\centerline{\includegraphics[angle=0,width=\textwidth]{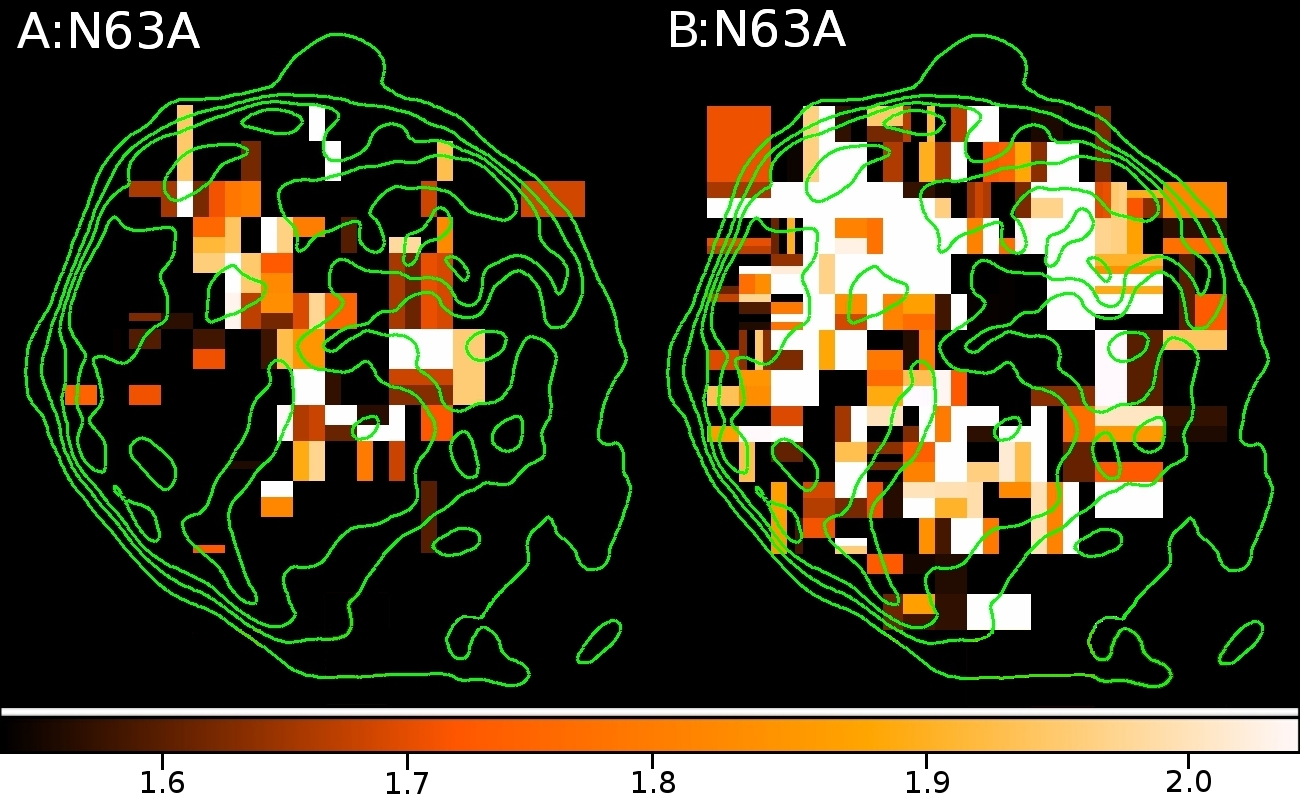}}
\figcaption[]{{$\chi_{\nu}^2$ maps of N63A based on our spectral model fits for 346 adaptive mesh-created regions (see Section 4.2). (A) The $\chi_{\nu}^2$ map for spectral model fits with abundances fixed at the mean ISM values of this work. (B) The $\chi_{\nu}^2$ map for spectral model fits with abundances fixed at the mean ISM values from H98.}\label{fig:fig7}}
\end{figure*}

\begin{figure*}[]
\figurenum{8}
\centerline{\includegraphics[angle=0,height=0.9\textheight,width=\textwidth]{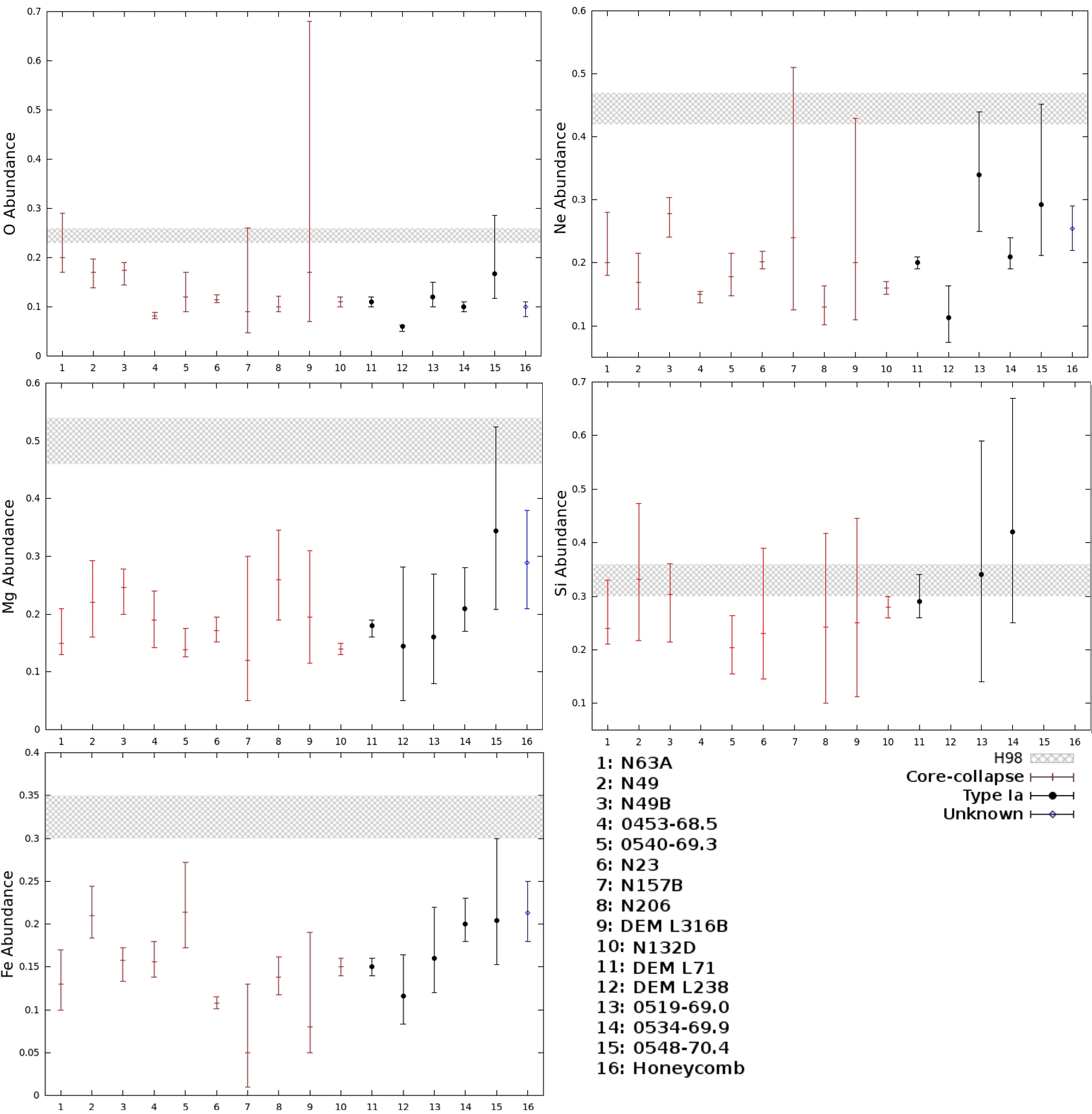}}
\figcaption[]{{Elemental abundances for O, Ne, Mg, Si, and Fe for each SNR used in this study. Abundances are with respect to solar (Anders and Grevesse 1989).}\label{fig:fig8}}
\end{figure*}

\begin{figure*}[]
\figurenum{9}
\centerline{\includegraphics[angle=0,width=\textwidth]{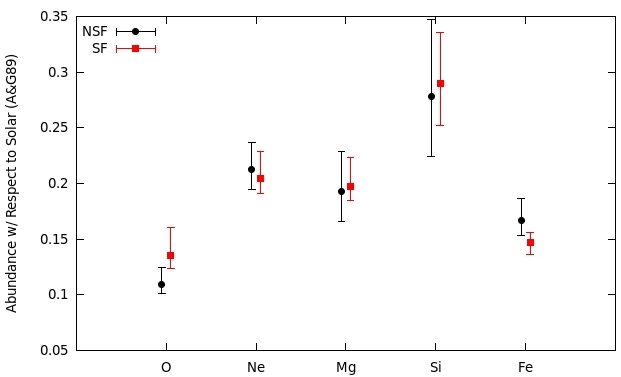}}
\figcaption[]{{Comparisons between SNRs located within star formation (SF) regions and those found in regions without star formation (NSF).}\label{fig:fig9}}
\end{figure*}

\begin{figure*}[]
\figurenum{10}
\centerline{\includegraphics[angle=0,width=\textwidth]{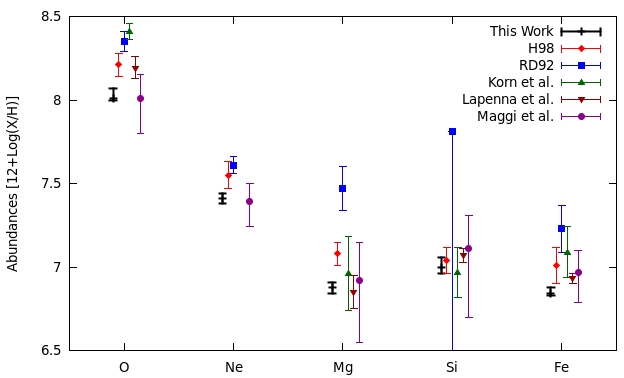}}
\figcaption[]{{Abundance comparisons among this work and previous measurements.}\label{fig:fig10}}
\end{figure*}

\begin{deluxetable*}{|l|c|c|c|c|c|c|}
\footnotesize
\tabletypesize{\scriptsize}
\tablecaption{Archival {\it Chandra} Data of LMC SNRs}
\tablewidth{0pt}
\tablehead{Target & ObsID(s) & RA & Dec & Observation Date(s) & Total Exposure & Refs\tablenotemark{c}\\ & & (J2000) & (J2000) & & (ks)&}
\startdata
0453--68.5 & 1990 & 04 53 40 & -68 29 45 & 2002-12-31 & 38 & 1\\
DEM L71 & 3876 4440 & 05 05 42 & -67 52 38 & 2003-07-04 2003-07-06 & 103 & 2\\
N23 & 2762 & 05 05 54 & -68 01 37 & 2004-01-07 & 37 & 3\\
0519--69.0 & 11241 12062 12063 & 05 19 35 & -69 02 08& 2009-12-25 2009-12-26 2010-02-03 & 51 & 4\\
N49B & 1041 & 05 25 25 & -65 59 22 & 2001-09-15 & 34 & 5\\
N132D & 5532 7259 7266 & 05 25 03 & -69 38 27 & 2006-01-09 2006-01-10 2006-01-15 & 89 & 6\\
N49 \tablenotemark{a} & 10123 10806 10807 10808 & 05 26 01 & -66 04 36 & 2009-07-18 2009-09-19 2009-16 2009-07-31 & 108 & 7\\
N206 & 3848 4421 &05 32 00 & -71 00 20 & 2003-02-27 2003-02-28 &  67 & 8\\
0534--69.9 & 1991 & 05 34 00 & -69 55 00 & 2001-09-01 & 59 & 9\\
DEM L238 & 3850 & 05 34 15 & -70 33 42 & 2003-10-02 & 67 & 10\\
N63A & 777 & 05 35 44 & -66 02 14 & 2000-10-16	& 43 & 11\\
& 1044 1967 2831 & & & 2001-04-25 2000-12-07 2001-12-12 & &\\
Honeycomb \tablenotemark{b}&  2832 3829  & 05 35 45 & -69 18 06 &  2002-05-15 2002-12-31  & 400 & 12\\
& 3830 4614 4615 & & & 2003-07-08 2004-01-02 2004-07-22 & &\\
N157B &	2783 & 05 37 48 & -69 10 20 & 2002-08-23 & 48 & 13\\
0540--69.3 & 5549 7270 7271 & 05 40 11 & -69 19 55 & 2006-02-15 2006-02-16 2006-02-18 & 104 & 14\\
DEM L316B & 2829 & 05 47 15 & -69 42 25 & 2002-07-27 & 35 & 15\\
0548--70.4 & 1992 & 05 47 50 & -70 24 45 & 2001-09-16 & 59 & 9\\
\enddata
\tablenotetext{a}{There are two other ObsIDs (1041 and 2515) that detected N49 in the {\it Chandra} archive. Those observations are not useful because of a large off-axis pointing ($\sim$6.5 for ObsID 1041) and a short exposure ($\sim$7 ks for ObsID 2515). Thus we excluded them in this work.}
\tablenotetext{b}{Honeycomb has not been the target object in any {\it Chandra} observations. However, Honeycomb is in close proximity to SNR 1987A ($\sim$1.7$\farcm$ south) and is thus detected on several SNR 1987A observations. We used 8 ObsIDs of SNR 1987A that include Honeycomb.}
\tablenotetext{c}{1. Gaensler et al. (2003); McEntaffer et al. (2012) 2. Hughes et al. (2003) 3. Hughes et al. (2006) 4. Hughes et al. (1995); Kosenko et al. (2010) 5. Park et al. (2003a) 6. Banas et al. (1997); Sano et al. (2015); Borkowski et al. (2007) 7. Park et al. (2003b); Park et al. (2012) 8. Williams et al. (2005) 9. Hendrick et al. (2003) 10. Borkowski et al. (2006a) 11. Warren et al. (2003) 12. Chu et al. (1995) 13. Chen et al. (2006) 14. Park et al. (2010) 15. Williams \& Chu (2005)}
\end{deluxetable*}
\clearpage

\begin{deluxetable*}{|l|c|c|cccccccc|l|}
\footnotesize
\tabletypesize{\scriptsize}
\tablecaption{Results of LMC ISM Abundance Measurements}
\tablehead{SNR & SNR & No. of & kT & $n_{e}$T & O & Ne & Mg & Si & Fe & $\chi_{\nu}^2$\tablenotemark{a} & SF\tablenotemark{b}\\
 	  Name & Type & Regions &keV& ($\times$ $10^{12}$ $cm^{-3}$s) & & & & & &}
\startdata
N63A & & 17 & $0.55^{+0.12}_{-0.02}$ & $3.6^{+3.3}_{-2.1}$ & $0.20^{+0.09}_{-0.03}$ & $0.20^{+0.08}_{-0.02}$ & $0.15^{+0.06}_{-0.02}$ & $0.24^{+0.09}_{-0.03}$ & $0.13^{+0.04}_{-0.02}$ & 1.1 & y\\
N49 & & 2 & $0.70^{+0.04}_{-0.05}$ & $2.1^{+3.5}_{-1.9}$ & $0.17^{+0.03}_{-0.03}$ & $0.17^{+0.05}_{-0.04}$ & $0.22^{+0.07}_{-0.06}$ & $0.33^{+0.14}_{-0.12}$ & $0.21^{+0.03}_{-0.03}$ & 1.0 & y\\
N49B & & 6 &$0.52^{+0.02}_{-0.05}$ & $2.1^{+3.1}_{-1.8}$ &  $0.17^{+0.02}_{-0.03}$ & $0.28^{+0.03}_{-0.04}$ & $0.25^{+0.03}_{-0.05}$ & $0.30^{+0.06}_{-0.09}$ & $0.16^{+0.01}_{-0.03}$ & 1.1 & y\\
0453-68.5 & & 6 & $0.35^{+0.02}_{-0.03}$ & $1.8^{+2.5}_{-2.3}$ &  $0.08^{+0.01}_{-0.01}$ & $0.15^{+0.01}_{-0.01}$ & $0.19^{+0.05}_{-0.05}$ & & $0.16^{+0.02}_{-0.02}$ & 1.2 & n\\
0540-69.3 & CC & 2 & $0.42^{+0.04}_{-0.05}$ & $2.9^{+1.8}_{-2.0}$ & $0.12^{+0.05}_{-0.03}$ & $0.18^{+0.04}_{-0.03}$ & $0.14^{+0.04}_{-0.01}$ & $0.20^{+0.06}_{-0.05}$ & $0.21^{+0.06}_{-0.04}$ & 1.1 & y\\
N23 & & 4 & $0.47^{+0.02}_{-0.03}$ & $0.4^{+1.2}_{-0.2}$ & $0.12^{+0.01}_{-0.01}$ & $0.20^{+0.02}_{-0.01}$ & $0.17^{+0.02}_{-0.02}$ & $0.23^{+0.16}_{-0.08}$ & $0.11^{+0.01}_{-0.01}$ & 1.0 & y\\
N157B & & 2 & $0.60^{+0.25}_{-0.30}$ & $0.6^{+2.0}_{-0.2}$ & $0.09^{+0.17}_{-0.04}$ & $0.24^{+0.27}_{-0.12}$ & $0.12^{+0.18}_{-0.07}$ & & $0.05^{+0.08}_{-0.04}$ & 1.2 & y\\
N206 & & 2 & $0.45^{+0.02}_{-0.01}$ & $2.0^{+3.7}_{-1.5}$ & $0.10^{+0.02}_{-0.01}$ & $0.13^{+0.03}_{-0.03}$ & $0.26^{+0.09}_{-0.07}$ & $0.24^{+0.18}_{-0.14}$ & $0.14^{+0.02}_{-0.02}$ & 1.1 & y\\
DEM L316B & & 1 & $0.49^{+0.10}_{-0.05}$ & $25.0^{+5.0}_{-9.1}$ & $0.17^{+0.51}_{-0.10}$ & $0.20^{+0.23}_{-0.09}$ & $0.20^{+0.12}_{-0.08}$ & $0.25^{+0.20}_{-0.14}$ & $0.08^{+0.11}_{-0.03}$ & 1.1 & y\\
N132D & & 8 & $0.73^{+0.02}_{-0.03}$ & $1.2^{+1.5}_{-1.0}$ & $0.11^{+0.01}_{-0.01}$ &	$0.16^{+0.01}_{-0.01}$ & $0.14^{+0.01}_{-0.01}$ & $0.28^{+0.02}_{-0.02}$ & $0.15^{+0.01}_{-0.01}$ & 1.3 & y\\\hline
DEM L71 & & 7 & $0.45^{+0.01}_{-0.01}$ & $2.7^{+3.4}_{-2.1}$ & $0.11^{+0.01}_{-0.01}$ & $0.20^{+0.01}_{-0.01}$ & $0.18^{+0.01}_{-0.02}$ & $0.29^{+0.05}_{-0.03}$ & $0.15^{+0.01}_{-0.01}$ & 1.2 & n\\
DEM L238 & & 2 & $0.44^{+0.04}_{-0.04}$ & $0.7^{+1.5}_{-0.5}$ & $0.06^{+0.01}_{-0.01}$ & $0.11^{+0.05}_{-0.04}$ & $0.15^{+0.14}_{-0.10}$ & & $0.12^{+0.05}_{-0.03}$ & 1.1 & n\\
0519-69.0 & Ia & 4 & $1.43^{+0.12}_{-0.15}$ & $0.5^{+0.8}_{-0.4}$ & $0.12^{+0.03}_{-0.02}$ & $0.34^{+0.10}_{-0.09}$ & $0.16^{+0.11}_{-0.08}$ & $0.34^{+0.25}_{-0.20}$ & $0.16^{+0.06}_{-0.04}$ & 1.2 & n\\
0534-69.9 & & 2 & $0.33^{+0.02}_{-0.02}$ & $3.2^{+5.5}_{-2.5}$ & $0.10^{+0.01}_{-0.01}$ & $0.21^{+0.03}_{-0.02}$ & $0.21^{+0.07}_{-0.04}$ & $0.42^{+0.25}_{-0.17}$ & $0.20^{+0.03}_{-0.02}$ & 1.2 & n\\
0548-70.4 & & 4 & $0.42^{+0.06}_{-0.04}$ & $2.6^{+2.0}_{-1.5}$ & $0.17^{+0.12}_{-0.05}$ & $0.29^{+0.16}_{-0.08}$ & $0.34^{+0.18}_{-0.14}$ & & $0.20^{+0.10}_{-0.05}$ & 1.2 & n\\\hline
Honeycomb & Unknown & 2 & $0.36^{+0.02}_{-0.01}$ & $1.1^{+1.3}_{-0.8}$ & $0.10^{+0.01}_{-0.02}$ & $0.25^{+0.04}_{-0.03}$ & $0.29^{+0.09}_{-0.08}$ & & $0.21^{+0.04}_{-0.03}$ & 1.1 & y\\\hline\hline
Mean & & & & & $0.13^{+0.01}_{-0.01}$ & $0.20^{+0.01}_{-0.01}$ & $0.20^{+0.02}_{-0.01}$ & $0.28^{+0.04}_{-0.03}$ & $0.15^{+0.01}_{-0.01}$ & & \\
$[$X/H$]$\tablenotemark{c} & & & & & $8.04\pm{0.04}$ & $7.39\pm{0.06}$ & $6.88\pm{0.06}$ & $6.99\pm{0.11}$ & $6.84\pm{0.05}$ & & \\ \hline
H98\tablenotemark{d} & & & & & $8.21\pm{0.07}$ & $7.55\pm{0.08}$ & $7.08\pm{0.07}$ & $7.04\pm{0.08}$ & $7.01\pm{0.11}$ & & \\
\enddata
\tablecomments{Abundances are relative to solar (Anders and Grevesse 1989). Statistical uncertainties are with 90\% confidence errors. For some SNRs the Si abundance is fixed at Russell \& Dopita (1992) value as we are unable to measure it due to poor photon statistics in the Si line feature.}
\tablenotetext{a}{The average $\chi_{\nu}^2$ of all spectral fits in the SNR.}
\tablenotetext{b}{y: SNRs in a star-forming environment. n: SNRs in a non-star-forming environment.}
\tablenotetext{c}{Mean abundances of this work expressed as [X/H] = 12 + Log(X/H)}
\tablenotetext{d}{Mean abundances of H98 expressed as [X/H] = 12 + Log(X/H)}
\end{deluxetable*}


\begin{thebibliography}{}

\bibitem[Anders \& Grevesse 1989]{ande89} Anders, E., \& Grevesse, N. 1989, Geochimica et Cosmochimica Acta, 53, 197

\bibitem[Banes et al. 1997]{banes97} Banas, K. R., Hughes, J. P., Bronfman, L., \& Nyman, L. 1997, ApJ, 480, 607

\bibitem[Badenes et al. 2006]{bad06} Badenes, C., Borkowski, K. J., Hughes, J.P., Hwang, \& U., Bravo, E. 2006, ApJ, 645, 1373

\bibitem[Badenes et al. 2009]{bad09} Badenes, C., Harris, J., Zaritsky, D., \& Prieto, J. L. 2009, ApJ, 700, 727

\bibitem[Borkowski et al. 2001]{bork01} Borkowski, K. J., Lyerly, W. J., \& Reynolds, S. P. 2001, ApJ, 548, 820

\bibitem[Borkowski et al. 2006a]{bork06a} Borkowski, K. J., Hendrick, S. P., \& Reynolds, S. P. 2006a, ApJ, 652, 1259

\bibitem[Borkowski et al. 2007]{bork07} Borkowski, K. J., Hendrick, S. P., \& Reynolds, S. P. 2007, ApJ, 671, L45

\bibitem[Chen et al. 2006]{chen06} Chen, Y., Wang, Q. D., Gotthelf, E. V., et al. 2006, ApJ, 651, 237

\bibitem[Chu \& Kennicutt]{CK88} Chu \& Kennicutt, R. C. 1988, AJ, 96, 1874

\bibitem[Chu et al. 1995]{chu95} Chu, Y.-H., Dickel, J. R., Staveley-Smith, L., Osterberg, J., \& Smith, R. C. 1995, AJ, 109, 1729

\bibitem[Cline et al. 1982]{cline82} Cline, T. L., Desai, U. D., Teegarden, B. J., et al. 1982, ApJ, 255, L45

\bibitem[Cole et al. 2005]{cole05} Cole, A. A., Tolstoy, E., Gallagher, III, J. S., \& Smecker-Hane, T. A. 2005, AJ, 129, 1465

\bibitem[Dickey \& Lockman 1990]{DL90} Dickey J.M., \& Lockman F.J., 1990, ARA\&A28, 215

\bibitem[Foster et al. 2012]{foster12} Foster, A. R., Ji, L., Smith, R. K., \& Brickhouse, N. S., 2012, ApJ, 756, 128

\bibitem[Gaensler et al. 2001]{gae01} Gaensler, B. M., Slane, P. O., Gotthelf, E. V., \& Vasisht, G. 2001, ApJ, 559, 963

\bibitem[Gaensler et al. 2003]{gae03} Gaensler, B. M., Hendrick, S. P., Reynolds, S. P., \& Borkowski, K. J. 2003, ApJL, 594, L111

\bibitem[Gotthelf \& Wang 2000]{gw00} Gotthelf, E. V., \& Wang, Q. D. 2000, ApJ, 532, L117

\bibitem[Hwang et al. 2001]{hwang01} Hwang, U., Petre, R., Holt, S. S., \& Szymkowiak, A. E. 2001, ApJ, 560, 742

\bibitem[Hendrick et al. 2003]{hen03} Hendrick, S. P., Borkowski, K. J., \& Reynolds, S. P. 2003, ApJ, 593, 370

\bibitem[Hughes et al. 1995]{H95} Hughes, J. P., et al. 1995, ApJ, 444, 81

\bibitem[Hughes et al. 1998]{H98} Hughes, J. P., Hayashi, I., \& Koyama, K. 1998, ApJ, 505, 732

\bibitem[Hughes et al. 2003]{H97} Hughes, J. P., Ghavamian, P., Rakowski, C. E., \& Slane, P. O. 2003, ApJ, 582, L95

\bibitem[Hughes et al. 2006]{H06} Hughes, J. P., Rafelski, M., Warren, J. S., et al. 2006, ApJ, 645, L117

\bibitem[Kaaret et al. 2001]{kaar01} Kaaret, P., et al. 2001, ApJ, 546, 1159

\bibitem[Klose et al. 2004]{klose04} Klose, S., Henden, A. A., Geppert, U., et al. 2004, ApJ, 609, L13

\bibitem[Korn et al. 2000]{korn00} Korn, A. J., Becker, S. R., Gummersbach, C. A., \& Wolf, B. 2000, A\&A, 353, 655

\bibitem[Kosenko, D., Helder, E. A., \& Vink, J. 2010]{Kos10} Kosenko, D., Helder, E. A., \& Vink, J. 2010, A\&A, 519, A11

\bibitem[Lapenna et al. 2012]{lap12} Lapenna, E., Mucciarelli, A., Origlia, L., \& Ferraro, F. R. 2012, ApJ, 761, 33

\bibitem[McConnachie 2012]{mcon2012} McConnachie, A. W. 2012, AJ, 144, 4

\bibitem[Maggie et al. 2015]{mag15} P. Maggi, F. Haberl, P. J. Kavanagh, et al. arXiv:1509.09223 [astro-ph.HE]

\bibitem[Manchester et al. 1993a]{man93a} Manchester, R. N., Mar, D. P., Lyne, A. G., Kaspi, V. M., \& Johnston, S. 1993a, ApJ, 403, L29

\bibitem[Manchester et al. 1993b]{man93b} Manchester, R. N., Staveley-Smith, L., \& Kesteven, M. J. 1993b, ApJ, 411, 756

\bibitem[McEntaffer et al. 2012]{McE12} McEntaffer, R. L., Brantseg, T., \& Presley, M. 2012, ApJ, 756, 17

\bibitem[Middleditch \& Pennypacker 1985]{mp85} Middleditch, J., \& Pennypacker, C. R. 1985, Nature, 313, 659

\bibitem[Morse, Winkler, \& Kirshner 1995]{M95} Morse, J. A., Winkler, P. F., \& Kirshner, R. P., 1995 AJ 109, 2104

\bibitem[Park et al. 2003a]{park03a} Park, S., Burrows, D. N., Garmire, G. P., et al. 2003a, ApJ, 586, 210

\bibitem[Park et al. 2003b]{park03b} Park, S., Hughes, J. P., Slane, P. O., et al. 2003b, ApJ, 592, L41

\bibitem[Park et al. 2010]{park2010} Park, S., Hughes, J. P., Slane, P. O., et al. 2010, ApJ, 710, 948

\bibitem[Park et al. 2012]{park12} Park, S., Hughes, J. P., Slane, P. O., et al. 2012, ApJ, 748, 117

\bibitem[Petre et al. 2007]{petre07} Petre, R., Hwang, U., Holt, S. S., Saﬁ-Harb, S., \& Williams, R. M. 2007, ApJ, 662, 988

\bibitem[Pompéia et al. 2009]{omp09} Pompéia, L., Hill, V., Spite, M., et al. 2008, A\&A, 480, 379

\bibitem[Rothschild et al. 1994]{roth94} Rothschild, R. E., Kulkarni, S. R., \& Lingenfelter, R. E. 1994, Nature, 368, 432

\bibitem[Russell \& Dopita 1992]{RD92} Russell, S. C., \& Dopita, M. A. 1992, ApJ, 384, 508

\bibitem[Smith et al. 2001]{smith01} Smith, R. K., Brickhouse, N.found S., Liedahl, D. A., \& Raymond, J. C. 2001, ApJ, 556, L91

\bibitem[Sonneborn et al. 1987]{sonn87} Sonneborn, G., Altner, B., \& Kirshner, R. P. 1987, ApJ, 323, L35

\bibitem[Warren et al. 2003]{warren03} Warren, J. S., Hughes, J. P., \& Slane, P. O. 2003, ApJ, 583, 260

\bibitem[Williams \& Chu 2005]{wc05} Williams, R. M., \& Chu, Y.-H. 2005, ApJ, 635, 1077

\bibitem[Williams et al. 2005]{will05} Williams, R. M., Chu, Y.-.H., Dickel, J. R., et al. 2005, ApJ, 628, 704

\bibitem[Van der Swaelmen et al. 2013]{van13} Van der Swaelmen, M., Hill, V., Primas, F., \& Cole, A. A. 2013, A\&A, 560, A44

\bibitem[Vancura et al. 1992]{van91} Vancura, O., Blare, W. P., Long, K. S., \& Raymond, J. C. 1992, ApJ, 394, 158


\end{thebibliography}
\end{document}